\documentclass[prl,twocolumn,showpacs]{revtex4}
\usepackage{amsfonts,amsmath,mathrsfs,epsfig,amsbsy,bm,verbatim}

\begin{document}
\title{Universal quantum computation in a semiconductor quantum wire network.}
\author{Jay D. Sau$^1$}
\author{Sumanta Tewari$^{2,1}$}
\author{S. Das Sarma$^1$}

\affiliation{$^1$Condensed Matter Theory Center and Joint Quantum Institute, Department of Physics, University of
Maryland, College Park, Maryland 20742-4111, USA\\
$^2$Department of Physics and Astronomy, Clemson University, Clemson, SC
29634}
\begin{abstract}
Universal quantum computation (UQC) using Majorana fermions on a 2D topological superconducting (TS) medium remains an outstanding open problem. This is because the quantum gate set that can be generated by braiding of the Majorana fermions does not include \emph{any} two-qubit gate and also the single-qubit $\pi/8$ phase gate. In principle, it is possible to create these crucial extra gates using quantum interference of Majorana fermion currents. However, it is not clear if the motion of the various order parameter
defects (vortices, domain walls, \emph{etc.}), to which the Majorana fermions are bound in a TS medium, can be quantum coherent. We show that these obstacles can be overcome using a semiconductor quantum wire network in the vicinity of an $s$-wave superconductor, by constructing topologically protected two-qubit 
gates and any arbitrary single-qubit phase gate in a topologically unprotected manner, which can be error corrected using magic state distillation.
Thus our strategy, using a judicious combination of topologically protected and unprotected gate operations,  realizes UQC on a quantum wire network with a remarkably high error threshold of $0.14$ as compared to $10^{-3}$ to $10^{-4}$ in ordinary 
unprotected quantum computation.

\end{abstract}

\maketitle

\paragraph{Introduction:}
The decoherence of quantum states by the environment is the nemesis of
any proposed quantum computation scheme. Topological quantum computation (TQC)
proposes~\cite{Kitaev,nayak_RevModPhys'08} an elegant way to solve this environmental decoherence problem
by encoding quantum information in an intrinsically non-local way. Quantum information thus
stored is expected to be essentially immune to any local perturbation due to the environment.
A class of quantum many-body states, characterized by excitations with non-Abelian statistics (non-Abelian anyons),
allow such non-local encoding of quantum information. In principle, the non-Abelian anyons can be moved (braided) around each other to exploit their statistics,
 which can be used to manipulate the stored quantum information and build quantum gates \cite{Parsa,Zhang,Freedman,Clarke}. Therefore, TQC using non-Abelian 
excitations is intrinsically fault-tolerant, which holds considerable promise to be able to beat the environmental decoherence problem.

Statistics \cite{Wilczek2} is defined as the unitary transformations on many-body wave functions
by the pair-wise exchange of the
particles' quantum numbers.
 In $(2+1)$-dimensions,
if the many-body ground state wave function happens to be a linear combination of states from a degenerate subspace,
a pair-wise exchange of the particles can unitarily \emph{rotate} the wave function in the ground state subspace.
In this case, the statistics is non-Abelian~\cite{Kitaev,nayak_RevModPhys'08} and the system of such quantum particles is a non-Abelian system.
Non-Abelian quantum systems in the so-called Ising topological class \cite{nayak_RevModPhys'08}
    are characterized by topological excitations called Majorana fermions.
In some topological superconducting (TS) systems \cite{schnyder}, Majorana fermions arise as non-degenerate zero-energy excitations bound to
 vortices of the superconducting order parameter. These topological excitations are protected
    from the higher-energy, non-topological, Bogoliubov excitations at the vortex cores \cite{Caroli} by the so-called
 mini-gap $\sim \frac{\Delta^2}{\epsilon_F}$, where $\Delta$ is the
      superconducting pair potential and $\epsilon_F$ is the Fermi energy.
      The second quantized operators, $\gamma_i$, corresponding to the Majorana excitations are self-hermitian, $\gamma_i^{\dagger}=\gamma_i$, which is in sharp contrast to ordinary fermionic (or bosonic) operators for which $c_i \neq c_i^{\dagger}$. Therefore, each Majorana particle is its own anti-particle \cite{Wilczek-3}
      unlike Dirac fermions where electrons and positrons (or holes) are distinct. Majorana particles have been predicted to occur in some exotic
 many-body states such as the proposed Pfaffian states in the filling fraction $\nu=5/2$ fractional quantum Hall (FQH) system \cite{Moore}, spin-less chiral $p$-wave superconductors/superfluids \cite{Read,Ivanov}, the surface of 3D strong topological insulators (TI) \cite{fu_prl'08}, and non-centrosymmetric superconductors \cite{Parag}.

    \paragraph{Semiconductor as a non-Abelian system:}  Recently, a semiconductor thin film with Rashba-type SO
    coupling was proposed to be a suitable platform for realizing a Majorana-carrying TS state by the proximity effect \cite{sau1,Ann}. It was shown that, in the presence of a $s$-wave superconducting pair potential $\Delta$ and a Zeeman splitting $V_z$, both of which can be proximity-induced ($V_z$ can also be induced
    by a parallel magnetic field when the SO coupling includes a Dresselhaus component \cite{alicea}), the appropriate TS state is realized when the parameters
    satisfy $V_z^2 > \Delta^2 + \mu^2$ where $\mu$ is the chemical potential in the semiconductor.
    Following this, it was quickly realized \cite{unpublished} that the 1D
version of the same set up, a semiconducting
 quantum wire with zero-energy Majorana states trapped at the
two ends,
 would be an easier system to explore the
physics of Majorana fermions, since the relevant mini-gap at the wire-ends is
of order $\Delta >> \frac{\Delta^2}{\epsilon_F}$  (there are no other sub-gap states localized near the ends other than
the Majorana states). It is important to note that $s$-wave proximity effect on a InAs quantum wire (which has a sizable SO coupling)
has possibly been already realized in experiments \cite{doh}. Moreover, the required Zeeman splitting $V$ in
the wire can be introduced more easily than in the 2D case by a magnetic field parallel
to the superconductor \cite{roman,Gil}. For all these reasons, it seems that a Majorana-carrying TS state in a semiconductor quantum wire may be within experimental reach. A discussion
of the SO coupled semiconductor as a non-Abelian platform in both 2D and 1D, along with STM signatures of Majorana modes from the wire-ends,
 can be found
in Ref.~[\onlinecite{long-PRB}].
\paragraph{Topological qubit using quantum wires:}
    Let us consider a semiconductor quantum wire in the TS state ($V^2 > \Delta^2 + \mu^2$). Each wire $i$ (shown as red segments in Fig. 1)
has a pair of Majorana modes $\gamma_i^{(L,R)}$ (shown as circles at the wire-ends) at the left (L) and right (R) ends. With wire $i$ we can associate a regular fermion state represented by the operator
$
d_i^\dagger=\frac{\gamma_i^{(L)}+\imath \gamma_i^{(R)}}{2}.
$
Thus, the wire naturally forms a two-state system consisting of states $|0\rangle$ and $|1\rangle = d_i^{\dagger}|0\rangle$, where $d_i|0\rangle =0$. Since the wave function for $d_i$ is composed of a pair of non-overlapping Majorana states, it is unaffected by all local changes in the Hamiltonian. Thus, the wire in the TS state constitutes a decoherence-free two state system which can be used to build a topologically protected qubit. However, such a two-state system does not allow the superposition of the basis states,
\emph{i.e}., the states ($|0\rangle \pm |1\rangle/\sqrt{2}$) do not exist, because they violate
the conservation of fermion parity \cite{bravyi1}. To
remedy this, a topological logical qubit can be defined~\cite{bravyi1}
via a \emph{pair} of quantum wires in the TS state, \emph{i.e.}, with the states $|\bar{0}\rangle=|00\rangle$ ($d$-states in both quantum wires unoccupied),
and $|\bar{1}\rangle=|11\rangle$ ($d$-states in both quantum wires occupied).
The superposition states, $(|\bar{0}\rangle\pm|\bar{1}\rangle)/\sqrt{2}$, are now allowed because the superconducting condenset only conserves fermion number modulo $2$.
Note also that these two states do not mix with the other two states $(|10\rangle, |01\rangle)$ of the two-wire system
by any unitary operation that conserves fermion parity.

   \paragraph{Quantum wire network and non-Abelian statistics:} Recently, a network of 1D semiconductor quantum wires has been proposed \cite{alicea1} as a suitable platform to create, transport, and fuse Majorana fermions at the wire-ends. The wire network consists of wire segments in the TS state (shown in red in Fig.~1) connected by segments in the non-topological superconducting (NTS) state (shown in blue in Fig.~1). The Majorana fermion states are transported by shifting the end points of the TS segments by applying locally tunable external gate potentials (which control $\mu$). That pair-wise exchange \cite{alicea1} of these Majorana fermions leads to the familiar
    non-Abelian statistics (\textit{i.e.,}
    $\gamma_i \rightarrow \gamma_j$ but $\gamma_j \rightarrow -\gamma_i$ ) follows most simply from  fermion parity conservation.
 Suppose $U$ is the unitary operator for exchange of Majorana fermions. Suppose also that $\gamma_i,\gamma_j$ do not
pick up a (relative) $-$ sign under $U$.  $U$ then transforms the neutral fermion operator $d^\dagger=\gamma_i+\imath\gamma_j$  into $Ud^\dagger U^\dagger=\imath d$. Applying $d^\dagger$ to $U|0\rangle$, where $|0\rangle$ is the empty state, it is easy to see that $U|0\rangle=\lambda|1\rangle=\lambda d^\dagger|0\rangle$, where $\lambda$ is a proportionality constant. This
contradicts fermion parity since $U$ is even and $d^\dagger$ is odd under fermion parity. It then follows that there must be a relative $-$ sign whenever two Majorana fermions are exchanged. Note also that the fact that the Majorana fermions in the present case are situated at the ends of 1D wires (and not in a 2D system like a 2D chiral $p$ wave superconductor) does not make any difference, since they are essentially zero-dimensional objects. In the wire network, these zero-dimensional objects are being moved (braided) on the 2D substrate of the superconductor. A more microscopic derivation of this non-Abelian statistics using Kitaev's 1D construction for a TS state \cite{Kitaev1} has been given in Ref.~[\onlinecite{alicea1}].

 The exchange and braiding operations on the Majorana fermions lead to some of the quantum gates such as the single-qubit $\pi/4$ phase gate and the single-qubit Hadamard gate. However, it is well known that \cite{bravyi1}, for a system of Majorana fermions, the exchange or braiding operations alone fail to provide any two-qubit gate: the topological
braiding operations allowed in a quantum wire network, as
in its 2D FQH or chiral $p$-wave Pfaffian counterpart, are not computationally sufficient. A system of
Majorana fermions can be made computationally sufficient if the
braiding-generated gate set is supplemented by a single-qubit $\pi/8$
phase gate and a two-qubit Controlled-Not, or CNOT, gate \cite{boykin,bravyi1}.

    \paragraph{Universal quantum computation with Majorana fermions:}
     A system of Majorana fermions can be made computationally sufficient in one of two ways \cite{Nayak,bravyi1}: (1) by dynamically changing the topology of
the platform, which allows the crucial extra gates to be obtained in a topologically
protected manner, or (2) by implementing these gates in
a topologically unprotected manner, which, provided the other gates are topologically 
protected, can also lead to universal quantum computation (UQC) with the aid of certain error correction protocols. At present, it is not clear how the topologically protected route
  can be implemented in any proposed TQC architecture including the quantum wire network. Therefore, in this paper, we take the second route to UQC as described above.
In the proposal we will consider, only the $\pi/8$ phase gate will be implemented in a topologically unprotected way. The topologically 
protected single qubit gates implemented through the braiding operations  can then be used to perform ``magic state distillation'' \cite{bravyikitaev,bravyi1}
 to produce error-corrected
$\pi/8$-phase gates from noisy ones. This purification protocol
(which has poly-log overhead) consumes several copies of a
magic state, e.g. $(|0\rangle + e^{i\pi/4} |1\rangle)/\sqrt{2}$, and outputs 
a single qubit with higher polarization along a magic direction.
Once a sufficiently pure magic state is produced, it may
then be consumed to generate a $\pi/8$-phase gate. This protocol
permits a remarkably high error threshold of over 0.14 for the
noisy gates,  as compared to $10^{-3}$ to $10^{-4}$ in ordinary 
unprotected quantum computation.
The simpler strategy of adopting the topologically unprotected route to UQC using a TS system still leaves us with a non-trivial problem.
The principal reason why any system of Majorana fermions is not computationally
sufficient is that two qubits cannot be entangled
using the braiding operations alone. Any logical
state of the two qubits, accessible by braiding one Majorana fermion around another, can always be written as a product of the
logical states of the individual qubits. It has been shown \cite{bravyi1} that a two-qubit CNOT gate can be created with Majorana fermions provided there is a supply of entangled pairs of two topological qubits. In the FQH context, it has been proposed \cite{bravyi1,Nayak,loss} that quantum interference of Majorana currents can be used to generate the two-qubit entanglement. However, in a TS system, in which the Majorana fermions are trapped in the order parameter defects (vortices or domain walls), it is not clear that the motion of these defects is a quantum process which will lead to the desired quantum interference. 
Further, creating a single-qubit $\pi/8$ phase gate in a TS system is also problematic. A simple method for this could be moving a pair of Majorana fermions in a given qubit near each other. Because of the overlap of the Majorana wave functions,
 the energy degeneracy of the $|0\rangle$ and the $|1\rangle$ states are then split. The dynamic phase in the resulting time evolution could then be used to produce arbitrary
single-qubit phase gates,
were it not for the fact that the Majorana wave functions in the TS medium are oscillatory in space, which results in corresponding oscillations in the energy splitting as a function of inter-Majorana distance \cite{meng}.
We show in this paper that both of these obstacles can be overcome in the quantum wire network, which therefore allows a concrete realization of a UQC architecture.

\begin{figure}[tbp]
\begin{center}
\includegraphics[width=0.4\textwidth]{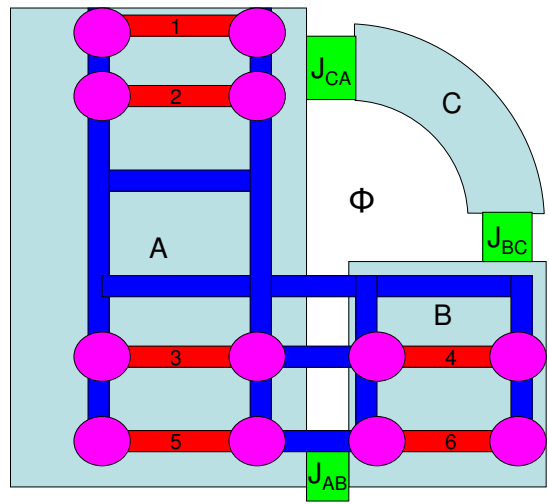}
\end{center}
\caption{(Color online) Schematic of entanglement generation in quantum wire
 topological qubits using superconductor Josephson junctions.  The
 light blue background represents superconducting islands labeled A, B and C.
 The dark blue segments are semiconductor quantum wires in the non-topological superconducting state,
while the red segments are semiconductor quantum wires in the topological superconducting state. The
purple circles represent the Majorana fermions at the end of topological
segments. The topological segments 1 and 2 form a representative topologically protected qubit.
Entanglement is generated in  the qubits formed by the topological segments $(3,4)$ and $(5,6)$
by transferring segments $4$ and $6$ to island B and then performing a
 quantum non-demolition
 measurement of the number of neutral fermions on island B using microwaves. $\Phi$ is the bias flux in the central hole.
  See text for details. }
\label{Fig1}
\end{figure}

\paragraph{UQC in the quantum wire network:}

 We first show that entangled pairs of qubits can be generated
by the set-up shown in Fig. \ref{Fig1}. The two superconducting islands B and C in Fig. \ref{Fig1},
together with the main superconductor A (which holds the wire network) constitute a three-island Josephson
junction flux qubit, which when biased with half a flux quantum,
has a degenerate pair of states composed of clock-wise supercurrent (CW)
and counter-clock-wise supercurrent (CCW). The charging energy of the
islands leads to tunneling between these two states leading to
a splitting of the degeneracy with the new eigenstates $\frac{1}{\sqrt{2}}$(CW$\pm$CCW). This splitting between the
energies is also sensitive to a Berry phase contribution which can be controlled by gate electrodes in the vicinity of the
islands \cite{tiwari}. In addition, as we show on the last page,
the splitting also depends on the parity of the number of neutral fermions
shared between the pairs of Majorana fermions at the ends of TS segments on island B (here we have assumed that the capacitance of the main island
A is large enough so we can ignore its charging energy). Using this, the system
can be tuned such that an even number of neutral fermions on island B leads to an exact degeneracy,
 while an odd number of neutral fermions leads to a
splitting between the states $\frac{1}{\sqrt{2}}$(CW$\pm$CCW). This
splitting can be measured by coupling the system to an rf circuit \cite{tiwari}, which can be used \cite{beenakker1} to perform a
non-demolition measurement of the state ($|0\rangle, |1\rangle$) of a pair of Majorana fermions on island B.
As has been emphasized in Ref.~[\onlinecite{beenakker1}], this provides an explicitly quantum mechanical method for the charge measurement
of a pair of Majorana fermions. Note that in the analogous method of charge measurement using
quantum interference of currents in a TS state, it is not clear if the motion of order parameter defects (vortices, domain walls) is a quantum
mechanical process.

  Let us now show how to use the quantum superposition states of the flux qubit to also create quantum entanglement between two topological qubits. The entangled state between the two qubits can then be used \cite{bravyi1} as the ancillary two-qubit states to construct a two-qubit quantum gate.
To generate an  entangled pair of qubits, we first create a pair of qubits composed of TS segments
 ($3$, $4$) and ($5$, $6$) both in the state $|\bar{0}\rangle\equiv|0,0\rangle$ on the main island A. By applying a Hadamard gate to both, we then transform the states of both qubits to $|\bar{0}\rangle+|\bar{1}\rangle$.
The combined state of the two-qubit system is now $(|\bar{0}\rangle+|\bar{1}\rangle)_{3,4}\otimes(|\bar{0}\rangle+|\bar{1}\rangle)_{5,6}$. We then transfer the TS segments $4$ and $6$ to island B by applying external gate potentials. If the parity of neutral fermions on segments $4$ and $6$ is even (odd), the degeneracy of the
states $\frac{1}{\sqrt{2}}$(CW$\pm$CCW) is split (not split). By an rf measurement, one can then collapse the quantum states of the two qubits as
\begin{equation}
(|\bar{0}\rangle+|\bar{1}\rangle)_{3,4}\otimes(|\bar{0}\rangle+|\bar{1}\rangle)_{5,6}\rightarrow |\bar{0}\rangle_{3,4}\otimes|\bar{0}\rangle_{5,6} + |\bar{1}\rangle_{3,4}\otimes|\bar{1}\rangle_{5,6},
\end{equation}
which is the desired entangled pair.
If in the rf measurement, no splitting is observed ($50 \%$ chance), the process has to be repeated until a splitting \emph{is} observed producing the desired entangled pair. Therefore, this method provides entangled pairs of qubits with a $50 \%$ success rate \emph{deterministically}.

In addition to two-qubit entanglement and a CNOT gate, for UQC, one needs a single-qubit $\pi/8$ phase gate.
As discussed earlier \cite{bravyi1,Nayak}, a simple way to create such a gate could be to bring  a pair of Majorana fermions from a topological qubit near each other and let the microscopic physics split the degeneracy between the $|0\rangle$ and $|1\rangle$ states. This has also been discussed in the context of general anyons.\cite{bonderson3} Any arbitrary single-qubit phase
  gate can then be created in principle by accumulating the relative dynamic phase between the $|\bar{0}\rangle$ and $|\bar{1}\rangle$ states over a finite period of time. However, it now appears that such a scheme does not work in both TS and FQH systems, because the splitting between the $|\bar{0}\rangle$ and $|\bar{1}\rangle$ states
oscillates with distance between the Majorana fermions in a given pair because of the oscillatory nature of the wave functions \cite{meng,Simon}. Recently an interferometric proposal has been suggested to 
avoid these oscillations in the FQH system. \cite{Clarke} At first glance, it appears that the same problem would arise in our quantum wire network, because the Majorana wave functions oscillate in the TS segments as a function of distance from the domain wall (oscillating solid black lines in Fig. 2a) just like in the case of a chiral $p$-wave superconductor. However, it is important to note that these functions \emph{do not oscillate} and in fact monotonically decay with a decay length inversely proportional to the gap (which is essentially proportional to the gate voltage $V_{gate}$ for $V_{gate} \gg \mu$) in the non-topological segments of the wire network (decaying solid black lines in Fig. 2a). Further, in the wire network, to induce splitting between the states $|\bar{0}\rangle$ and $|\bar{1}\rangle$ between a pair of Majorana end states, one does not need to physically move these states near each other (a process which is prone to errors). Instead, one could simply reduce $V_{gate}$ in the non-topological segments to generate the required overlap of the Majorana fermion wave functions on the two ends.

The presence of oscillations in the wave-functions in the TS and the 
absence thereof in the NTS can be understood from simple considerations 
of the asymptotic wave-functions. The TS segment consists of a superconducting state on a Fermi-liquid like state with Fermi wave-vector $k_F$. 
Therefore the wave-function of a zero-energy Majorana state has an 
asymptotic form $\psi(r)\propto e^{-r/\xi}e^{i k_F r}$ where the 
coherence length $\xi=\Delta/v_F$.\cite{long-PRB}
 The overlaps of this wave-function $\psi$ clearly has oscillations at 
the wave-vector $k_F$ as is the case with $p$-wave superconductors.\cite{meng} In contrast the NTS part of the system is depleted and therefore has 
a vanishing Fermi-wave-vector $(k_F=0)$. Thus the overlaps across 
the NTS has a purely decaying form as is clear from Fig.~2.
 Below, with the help of Fig.~2,
 we explicitly show how an arbitrary single-qubit gate can be constructed.

 From the plot of the wave function (solid black line) in Fig. 2(a), it is clear that the wave functions
no longer oscillate in the NTS segment. To generate a phase-shift, we first distort the TS segment 1 in the qubit in Fig.~2(b) into a
U shape. This step is necessary to bring the two Majorana modes at the two ends of segment 1 separated by a NTS region.  The gate voltage $V_{gate}$ in the NTS segment is still kept large such that the overlap of the wave functions is still negligible.
 We now lower the gate voltage in the NTS segment to increase the decay lengths of the wave functions. This causes overlap between the Majorana fermion wave functions at the ends of the NTS segment.
 As seen in Fig 2(b), this
 leads to a clear non-oscillatory dependence of the energy splitting between the
$|\bar{0}\rangle$ and $|\bar{1}\rangle$ states as a function of the applied gate voltage.

\begin{figure}[tbp]
\begin{center}
\includegraphics[width=0.35\textwidth]{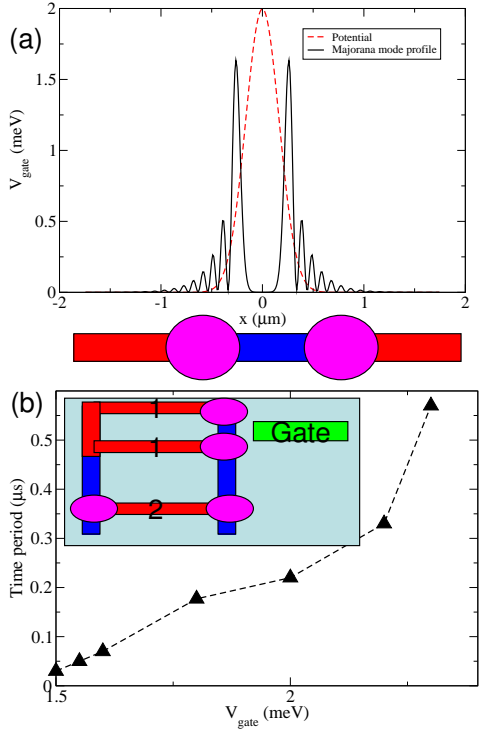}
\end{center}
\caption{(Color online)(a) Majorana wave-functions (black lines)
are fused across a non-topological wire segment (blue segment) with a gaussian
potential (red dashed line) with
peak height 2.0 meV for an InAs nanowire. Note that there are no oscillations of the wave functions in the non-topological blue segment. The
wave function, on the other hand, rapidly oscillates in the red topological segments. The purely decaying wave-functions
in the non-topological segment lead to a well-controlled phase shift as a function of time.
(b) Inset shows arbitrary single-qubit phase gate constructed by controlled
overlap of Majorana fermion wave functions through the non-topological segment. The plot shows the dependence of the
 phase rotation period on the barrier height. }
\label{Fig2}
\end{figure}

A $\pi/8$ phase gate is obtained by applying a $V_{gate}$ pulse over an appropriate
length of time. The appropriate length of time can be estimated by \emph{experimentally} determining the energy splitting between the $|\bar{0}\rangle$ and $|\bar{1}\rangle$ states. The rate of phase shift $\Delta E$ can
be determined by creating a test state $|\bar{0}\rangle$ and applying the sequence (Hadamard $\rightarrow$ phase gate $\rightarrow$ inverse Hadamard) in order. After this process, the probability
 of making a transition to the  $|\bar{1}\rangle$ state is given by
\begin{equation}
P(t)=\frac{1-\cos{\Delta E t}}{2}
\end{equation}
By experimentally determining this probability at time $t$, one gets $\Delta E$ for a given $t$ and $V_{gate}$. Arbitrary single-qubit phase
gates can then be constructed by calculating the required time and accumulating the correct phase shift for a given $V_{gate}$.

\paragraph{Majorana Berry phase in the flux qubit:}
To understand the origin of the Berry phase term generated by Majorana fermions it is necessary to
revert back to the fermion description of the system. This description is in terms of the fermion fields $\psi_\alpha$
for the electrons in both the semiconducting wires and on the  superconducting island $\alpha=A,B,C$.
 The pairing interactions $g_\alpha$ leading to Cooper-pairing on the islands together with Coulomb
interaction induced capacitance
 $C_\alpha$ and gate potential $V^{ext}_\alpha$ can be described in terms of an action
\begin{align}
&S=\int \imath \bar{\psi}_\alpha\partial_t\psi_\alpha -H_\alpha[ \bar{\psi}_\alpha,\psi_\alpha]-(V_\alpha+V^{ext}_\alpha)\bar{\psi}_\alpha\psi_\alpha\nonumber\\
&-(\Delta_\alpha\bar{\psi}_\alpha\bar{\psi}_\alpha+\Delta^*_\alpha\psi_\alpha\psi_\alpha)+C_\alpha V_\alpha^2/2+|\Delta_\alpha^2|/2g_\alpha+t_{\alpha\beta}\bar{\psi}_\alpha\psi_\beta
\end{align}
where $V_\alpha$ and $\Delta_\alpha$ are the effective Hubbard-Stratonovich fields which may be interpreted as
the time-dependent mean-field electrostatic potential and pairing potential respectively. The relatively
 weak tunnelings $t_{\alpha\beta}$
 between the islands and also the wires on the different islands will give rise to the
Josephson couplings between the islands. For our calculation, we will consider a phase model
such that $|\Delta_\alpha|=|\Delta|$. We will also assume that the capacitances of the islands is large enough so that
we can assume the quantum fluctuations in $V_\alpha$ and $\phi_\alpha$ to be slowly varying. In this limit,
 the fermionic part of the fields that contribute to the partition function  may be  assumed to follow
ground-state of the Hamiltonian
\begin{align}
&H_{mf}[V_\alpha,\phi_\alpha]=H_\alpha[ \bar{\psi}_\alpha,\psi_\alpha]+(V_\alpha+V^{ext}_\alpha)\bar{\psi}_\alpha\psi_\alpha\nonumber\\
&+(\Delta_\alpha\bar{\psi}_\alpha\bar{\psi}_\alpha+\Delta^*_\alpha\psi_\alpha\psi_\alpha)+t_{\alpha\beta}\bar{\psi}_\alpha\psi_\beta
\end{align}
as $\phi_\alpha$ and $V_\alpha$ vary in time adiabatically. For systems containing Majorana fermions the ground state is degenerate. The Majorana fermion
sector of the Hamiltonian may be described in terms of pairs of Majorana fermions combined into regular zero energy fermions $d^\dagger_{m\alpha}$ that
are localized on the island $\alpha$. The relevant states may then be charcterized by $Q^\dagger|\phi_\alpha\rangle$ where $|\phi_\alpha\rangle$
denotes the ground state with all Majorana fermion states empty and $Q^\dagger=\prod d^{\dagger n_{m\alpha}}_{m\alpha}$ is the operator that accounts for the
Majorana state occupation.
The transition matrix element between the various phase states $Q_{i}^\dagger|\phi_{\alpha,i}\rangle$ and $Q_{f}^\dagger|\phi_{\alpha,f}\rangle$ is given by
\begin{equation}
\mathcal{T}=\sum_{w_\alpha} \int_{\phi_{\alpha,i}}^{\phi_{\alpha,f}+2w_\alpha\pi}\mathcal{D}\phi_\alpha \langle \phi_{\alpha,f}| Q_{f} U(t_f,t_i) Q_{i}^\dagger|\phi_{\alpha,i}\rangle
\end{equation}
where $d U(t,t_i)/dt = H_{mf}(t)U(t,t_i)$ is the unitary time evolution matrix for the fermionic state over a particular phase trajectory. The states $Q_{i}^\dagger|\phi_{\alpha,i}\rangle$ and $Q_{f}^\dagger|\phi_{\alpha,f}\rangle$ are groundstates of $H_{mf}(t_i)$ and $H_{mf}(t_f)$ respectively with appropriate Majorana state occupancy.

Since the zero-energy fermion operators $d_{m\alpha}^\dagger$ are spatially separated, and localized on each  island $\alpha$, they evolve with phase $\phi_\alpha$ according to
\begin{equation}
d_{m\alpha}^\dagger=\int dr u_{m\alpha}(r)\psi^\dagger_\alpha(r)e^{\imath \int \dot{\phi}_\alpha/2}+ u_{m\alpha}(r)\psi_\alpha(r)e^{-\imath \int \dot{\phi}_\alpha/2}.
\end{equation}
Moreover, because of the absence of tunneling between the Majorana fermions, the occupation of each of these zero energy modes is conserved during the evolution
of the Hamiltonian. Therefore under the time-evolution $U(t_f,t_i)$, $Q_{i}^\dagger$ evolves into $U(t_f,t_i)Q_{i}U(t_f,t_i)^\dagger=Q_{f}^\dagger(-1)^{\sum_{\alpha}w_\alpha n_\alpha}$ where $n_\alpha=\sum_m n_{m\alpha}$ and $w_\alpha$ is the winding number of the phase trajectory along a particular phase $\phi_\alpha$. Furthermore $Q_{f}Q_{f}^\dagger |\phi_{\alpha,f}\rangle= |\phi_{\alpha,f}\rangle$. Therefore the transition amplitude for a given fermion occupation on each island $n_\alpha$ is
\begin{equation}
\mathcal{T}(n_\alpha)=\sum_{w_\alpha} \int_{\phi_{\alpha,i}}^{\phi_{\alpha,f}+2w_\alpha\pi}\!\!\mathcal{D}\phi_\alpha (-1)^{w_\alpha n_\alpha}\langle \phi_{\alpha,f}| U(t_f,t_i)|\phi_{\alpha,i}\rangle.
\end{equation}
The factor $(-1)^{\sum_{\alpha}w_\alpha n_\alpha}$ is precisely the Berry phase term associated with Majorana fermions and can be accounted for
by adding a term $\dot{\phi}_\alpha n_\alpha/2$ to the phase action of the system. Adding this to the phase action
 considered by Tiwari and Stroud \cite{tiwari}, which can be obtained by performing the $V_\alpha$ integral, is
\begin{align}
&S[n_{\alpha},\phi_\alpha]\approx \dot{\phi}^2_\alpha/2C_\alpha+\dot{\phi}_\alpha[\frac{V^{ext}_\alpha}{C_\alpha}+n_{\alpha}]-J_{\alpha\beta}\cos{(\phi_\alpha-\phi_\beta)}.
\end{align}

In the presence of an externally applied flux $\Phi$ as shown in Fig. 1, we replace the phase differences in the above equation
by the gauge-invariant phase differences $\phi_1=\phi_B-\phi_A\rightarrow \phi+a_1$,  $\phi_C-\phi_B\rightarrow (-\phi+\phi'+a_2)$
and $\phi_2=\phi_A-\phi_C\rightarrow -\phi'+a_3$. The Josephson couplings are taken to be $J_{AB}=J_{CA}=J$ and $J_{BC}=\alpha J$.
The gauge potentials are chosen such that  $a_1=a_2=a_3=2\pi\Phi/(3 \Phi_0)$. Here the phase of the island $B$, $\phi_B=\phi$ and $\phi_C=\phi'$.

The lowest Josephson energy configurations of this system are given by
$(\phi_1,\phi_2)=(\phi^*+2m\pi,-\phi^*+2n\pi),(-\phi^*+2m\pi,\phi^*+2n\pi)$
where $\phi^*=\cos^{-1}{\frac{1}{2\alpha}}$.\cite {tiwari}
In the geometry considered, ignoring the large capacitances $C_A^{-1}=C_C^{-1}=0$,
 the capacitance term $\dot{\phi}^2/2 C$ can
cause tunneling between the 2 minima.\cite{averin}
For the case $\alpha>1$, and starting at $(\phi^*,-\phi^*)$, there are 2
equivalent in energy low barrier tunneling paths to 2 equivalent
points $(2\pi-\phi^*,\phi^*)$ and $(-\phi^*,-2\pi+\phi^*)$ which
are related to each other by the symmetry $\phi_1\leftrightarrow-\phi_2$.

The total tunneling matrix element using the instanton approach\cite{averin}, is now given by the
imaginary time action $S_{E,j}$ for tunneling path $j$ as
\begin{equation}
\Gamma=\sum_{j=1,2} \omega_j e^{-S_{E,j}}
\end{equation}
where $\omega_j$ is the attempt frequency.
The contributions to the tunneling amplitudes from the 2 paths are identical by symmetry except for the term $\dot{\phi}(V^{ext}_B-\bar{d}_{mB}d_{mB}/2)$ which
creates a difference in the 2 paths of
\begin{equation}
S_{E,1}-S_{E,2}=2\imath\pi (Q-n_B/2)
\end{equation}
where $n_B=\sum_m \bar{d}_{mB}d_{mB}$ is the total number of fermions on island $B$.
This leads to an interferometric dependence
\begin{equation}
|\Gamma|=\sqrt{2}\Gamma_0 \cos{\pi(Q-n_B/2)}
\end{equation}
and $\Gamma_0$ is the single path tunneling amplitude.
The interference effect between the 2 paths may be interpreted as a
flux tunneling around the charge and has been referred to as a
Aharanov-Casher effect \cite{averin}.
Tuning $Q=0$, this leads to a splitting for $n_0$ (mod 2)=1 and no splitting for
$n_0$ (mod 2)=0.

The magnitude of the splitting $\Delta$, which contains information
about the Aharanov-Casher phase, can be measured by
 applying a flux of $\Phi=\Phi_0/2$. In this case the two minimum
Josephson energy configurations are degenerate and any splitting that is
measured is a result of the Aharanov Casher phase.

\paragraph{Conclusion:} We show that a network of semiconductor quantum wires in the vicinity of
 an $s$-wave superconductor
 allows universal quantum computation. To do this, we propose a scheme to generate
entanglement between two topological qubits in
the wire network with the assistance of a
flux qubit. We also show that the wave functions of the Majorana fermion states at the end points of the topological segments \emph{do not oscillate}
in the adjoining non-topological segments, even though they have the familiar oscillatory behavior~\cite{meng,Simon} in the topological segments. This fact can be used to create arbitrary single-qubit phase gates by controlled overlap of the Majorana wave functions \emph{via the non-topological segments} of the wire network. Our schemes for deterministically generating two-qubit entanglement and arbitrary single-qubit phase gates establish the
semiconductor wire network as a viable platform for universal quantum computation.


This work is supported by DARPA-QuEST, JQI-NSF-PFC, and LPS-NSA. We thank the Aspen Center for Physics where a part of this work was completed. ST acknowledges DOE/EPSCoR Grant \# DE-FG02-04ER-46139 and Clemson University start up funds for support.


\end{document}